\documentclass[%
 reprint,
 amsmath,amssymb,
 aps,
]{revtex4-1}
\usepackage{graphicx}
\usepackage{color}
\usepackage{amsmath}
\usepackage{amssymb}
\usepackage{amsthm}
\usepackage{booktabs}
\usepackage{float}
\usepackage[dvipdfm,colorlinks,urlcolor=blue,linkcolor=blue,anchorcolor=blue,citecolor=blue]{hyperref}

\begin{document}

\title{Effect of pressure on thermalization of one-dimensional nonlinear chains}

\author{Weicheng Fu$^{1,3}$}
\author{Yong Zhang$^2$}
\email{yzhang75@xmu.edu.cn}
\author{Hong Zhao$^{2,3}$}
\affiliation{$^1$Department of Physics, Tianshui Normal University, Tianshui 741001, Gansu, China\\
$^2$Department of Physics, Xiamen University, Xiamen 361005, Fujian, China\\
$^3$Lanzhou Center for Theoretical Physics, Key Laboratory of Theoretical Physics of Gansu Province, Lanzhou University, Lanzhou, Gansu 730000, China}

\date{\today }

\begin{abstract}
Pressure plays a vital role in changing the transport properties of matter. To understand this phenomenon at a microscopic level, we here focus on a more fundamental problem, i.e., how pressure affects the thermalization properties of solids. As illustrating examples, we study the thermalization behavior of the monatomic chain and the mass-disordered chain of Fermi-Pasta-Ulam-Tsingou-$\beta$ under different strains in the thermodynamic limit. It is found that the pressure-induced change in nonintegrability results in qualitatively different thermalization processes for the two kinds of chains. However, for both cases, the thermalization time follows the same law --- it is inversely proportional to the square of the nonintegrability strength. This result suggests that pressure can significantly change the integrability of a system, which provides a new perspective for understanding the pressure-dependent thermal transport behavior.
\end{abstract}

\maketitle

\textit{Introduction}.---Pressure can strongly affect the properties of phonons and thus change the phonon-related transport behavior of solids. High-temperature superconductivity of hydrogen-rich materials (e.g., H$_3$S and LaH$_{10}$) at high pressure is an outstanding example~\cite{RN334,RN335,PhysRevB.99.140501,RN336,
FLORESLIVAS20201,Pickard2020}. Besides, the heat conduction of some semiconductors~\cite{RN337} and low-dimensional materials~\cite{PhysRevB.79.205433,
PhysRevB.81.245318,C8CP06414H,RN338} has complex dependence on pressure, which paves a new way for thermal engineering. Even for simple one-dimensional (1D) nonlinear chains, it has been observed that pressure can qualitatively modify the transport behavior of phonons~\cite{PhysRevB.92.245411}. More interesting is the pressure-driven crossover from anomalous to normal heat conduction~\cite{Jiang_2016,PhysRevE.94.012115,PhysRevE.102.012111}. To understand the pressure-dependent transport behavior at a microscopic level, here we intend to investigate a more fundamental problem, i.e., how pressure affects the thermalization properties of solids.

The origin of the study of thermalization can be traced back to the famous ergodic hypothesis, which was formulated by Boltzmann in the 1870s, and is at the foundation of statistical physics~\cite{1949Khinchin}. This hypothesis leads to a main result that is the equivalence of time averages with phase averages, which implies the equipartition of energy among the various degrees of freedom. More than half a century later, in the 1950s, Fermi led Pasta, Ulam, and Tsingou (FPUT)~\cite{Fermi1955,dauxois:ensl-00202296} to conduct the first computer simulation and to verify this hypothesis by observing the rates of mixing and thermalization in a microscopic reversible dynamical system. However, the result was contrary to general expectations, i.e., the simulation failed to find the energy equipartition but discovered the celebrated ``FPUT \textit{recurrences}'' (or called ``FPUT \textit{paradox}''). This seminal work stimulated a substantial amount of research on the subject~\cite{Chaos2005,2008LNP728G} (see references therein) and revitalized the two research fields: nonlinear science and computational science~\cite{doi:10.1063/1.1861554,doi:10.1063/1.1889345,Porter2009Fermi}. However, in general, such a nonlinear many-body problem is analytically unsolvable. Thus, to answer whether a generic Hamiltonian system that far away from the equilibrium will eventually enter the thermalized state is still one of the most challenging problems in statistical physics~\cite{2008LNP728G}.

Recently, the wave turbulence approach was applied to attack the problem of thermalization~\cite{Onorato4208,PhysRevLett.120.144301,0295-5075-121-4-44003}. It is suggested that the thermalization time, $T_\text{eq}$, of a short FPUT chain follows a power law, which leads to a conclusion that the thermalized state will be reached for arbitrary small nonlinearity~\cite{Onorato4208}. More recently, we have found, via extensive numerical simulations, that the thermalization behavior of 1D near-integrable systems exhibit universality in the thermodynamic limit~\cite{Our2018,Fu_2019,PhysRevE.100.052102,PhysRevLett.124.186401}. It is shown that if the perturbation strength, $\epsilon$, of a system is defined accurately by selecting a suitable reference integrable system, $T_\text{eq}$ follows a universal scaling law, i.e., $T_\text{eq}\propto \epsilon^{-2}$. In particular, the results of diatomic chains show that different ways of perturbation may result in the qualitative difference in the processes of thermalization, but the scaling of $T_\text{eq}$ follows the same rule~\cite{PhysRevE.100.052102}. It is noted that the thermalization in the Klein-Gordon lattice~\cite{PhysRevE.94.062104,Pistone2018} also follows this law in the weakly nonlinear region, though this lattice belongs to another class that possesses on-site potential. Subsequent studies show that this universal scaling law is even independent of the dimension of a system, e.g., $T_\text{eq}$ of two- and three-dimensional systems follow the same law of scaling~\cite{2020arXiv200503478W}.

However, in almost all the studies on thermalization problems, there is little work on how pressure affects the thermalization properties of a system. Thus, we here take the 1D FPUT-$\beta$ chains in the two cases of monatomic and mass-disordered under strain as examples to study the effect of pressure on thermalization. It is hoped that this study could shed some light on understanding the micro-mechanism of relaxation and transport properties varying with pressure.

\textit{The models}.---We consider a chain consisting of $N+2$ particles with fixed ends, its Hamiltonian is
\begin{equation}\label{eqHam}
  H=\sum_{i=1}^N\frac{p_i^2}{2m_i}+\sum_{i=0}^NV(x_{i+1}-x_{i}-a),
\end{equation}
where $p_i$, $m_i$ and $x_i$ are, respectively, the momentum, mass, and position of the $i$th particle, and $a$ is the lattice constant, i.e., the average distance between adjacent particles in the natural state, and $x_0=0,~x_{N+1}=(N+1)a$. $V$ is the nearest-neighboring interaction potential, and it takes the form of FPUT-$\beta$: $V(x)=x^2/2+x^4/4$. Here, we studied two cases: the monatomic and the mass-disordered. For the former, the masses $m_i=1$; for the latter, the mass is set to be a random number uniformly distributed in $[1-\Delta m,1+\Delta m]$, where $\Delta m$ is the strength of disorder, and the mean value of masses is fixed to be unity.

Positive and negative pressure can be exerted by reducing or increasing the length of the chain, respectively. Let $b$ denote the new equilibrium distance between adjacent particles when the chain is stretched, so the strain $l=b-a$, and $l>0$ (or $l<0$) means the chain is tensile (or compressive). Therefore, the potential between adjacent particles in the strained chain becomes
\begin{equation}\label{eqV}
\begin{aligned}
  &V(x+l)=\frac{(x+l)^2}{2}+\frac{(x+l)^4}{4}=\\
  &\frac{\left(2l^2+l^4\right)}{4}+\left(l+l^3\right) x
  +\frac{\left(3 l^2+1\right)}{2} x^2+l x^3+\frac{x^4}{4}.
\end{aligned}
\end{equation}
Note that the first term at the right-hand side is a shift of the potential, and it does not enter the motion equations of the system. The second term, which represents a constant force acting on a particle, is also trivial for the dynamics of the chain. Because each particle is subjected to a pair of forces of equal magnitude and opposite direction, which will cancel each other. Consequently, the last three terms rule the motions of the system.

From Eq.~(\ref{eqV}), it is also seen that the variation of the strain will change the strength of the harmonic term and the third-order nonlinear term simultaneously. However, the change of the harmonic term means that the eigenfrequencies (characteristic time) of the system will be changed. To compare the rate of thermalization (RT) of the system under different strains, the timescale needs to keep same. Given this, the interaction potential~(\ref{eqV}) is rescaled as
\begin{equation}\label{eqV2}
  \tilde{V}(x+l)=C+\frac{1}{2} x^2+\frac{\lambda_3}{3} x^3+\frac{\lambda_4 x^4}{4},
\end{equation}
where $C=\left[\left(2l^2+l^4\right)/4+\left(l+l^3\right) x\right]({3 l^2+1})^{-1}$, and $\lambda_3={3l}({3 l^2+1})^{-1}$, and $\lambda_4=(3 l^2+1)^{-1}$. Namely, the stretched FPUT-$\beta$ chain becomes the FPUT-$\alpha$-$\beta$ one with strain-dependent nonlinear coefficients.

To facilitate perturbation analysis, the Hamiltonian of the system is generally written as
\begin{equation}
  H=H_0+H',
\end{equation}
where $H_0$ and $H'$ denote the integrable part and the perturbation, respectively. The form and the strength of $H'$ depend on the choice of $H_0$. A conventional practice is to take the Hamiltonian of the \emph{harmonic} system as $H_0$ and defines the rest nonlinear part as the perturbation, that is, the anharmonicity strength is the perturbation strength. Thus, for Eq.~(\ref{eqV2}), the perturbation strength can be given as a dimensionless form
\begin{equation}\label{eqPH}
    \epsilon_{\text{H},n}=|\lambda_n|\varepsilon^{n/2-1},\quad n=3,~4,
\end{equation}
where $n$ is the order of perturbation, and $\varepsilon$ is the energy per particle. The dynamics for this choice of $H_0$ is denoted as perturbed harmonic dynamics (PHD).

However, the anharmonicity does not necessarily destroy the integrability of the system. For example, the Toda model~\cite{1967Toda} is anharmonic but integrable. Thus, the Hamiltonian of the \emph{Toda} as $H_0$ is an alternative, and the corresponding dynamics is denoted as perturbed Toda dynamics (PTD). Some research shows that it is a better choice to treat a generic monatomic nonlinear chain as the Toda's perturbation~\cite{FERGUSON1982157,PhysRevE.55.6566,PhysRevE.62.6078,Benettin2011,Benettin2013,Benettin2018,Fu_2019}. Such as the chain described by Eq.~(\ref{eqV2}) can be regarded as the Toda potential $V_\text{T}(x)=\frac{e^{2\lambda_3x}-2\lambda_3x-1}{4\lambda_3^2}$ plus the perturbation~\cite{Fu_2019}
\begin{equation}
  H'=-\sum_{n=4}^\infty \frac{\delta_n}{n}x^n,
\end{equation}
where the coefficient
\begin{equation}\label{eqDn}
\delta_n=
  \begin{cases}
    \frac{3l^2-1}{(3l^2+1)^2},&n=4,\\
    \frac{1}{(n-1)!}\left(\frac{6l}{3l^2+1}\right)^{n-2},&n\geq5,
  \end{cases}
\end{equation}
and the dimensionless perturbation strength is
\begin{equation}\label{eqPT}
    \epsilon_{\text{T},n}=|\delta_n|\varepsilon^{n/2-1},\quad n=4,~5,~\cdots,\infty.
\end{equation}

\begin{figure}[t]
	\centering
    \includegraphics[width=0.5\textwidth]{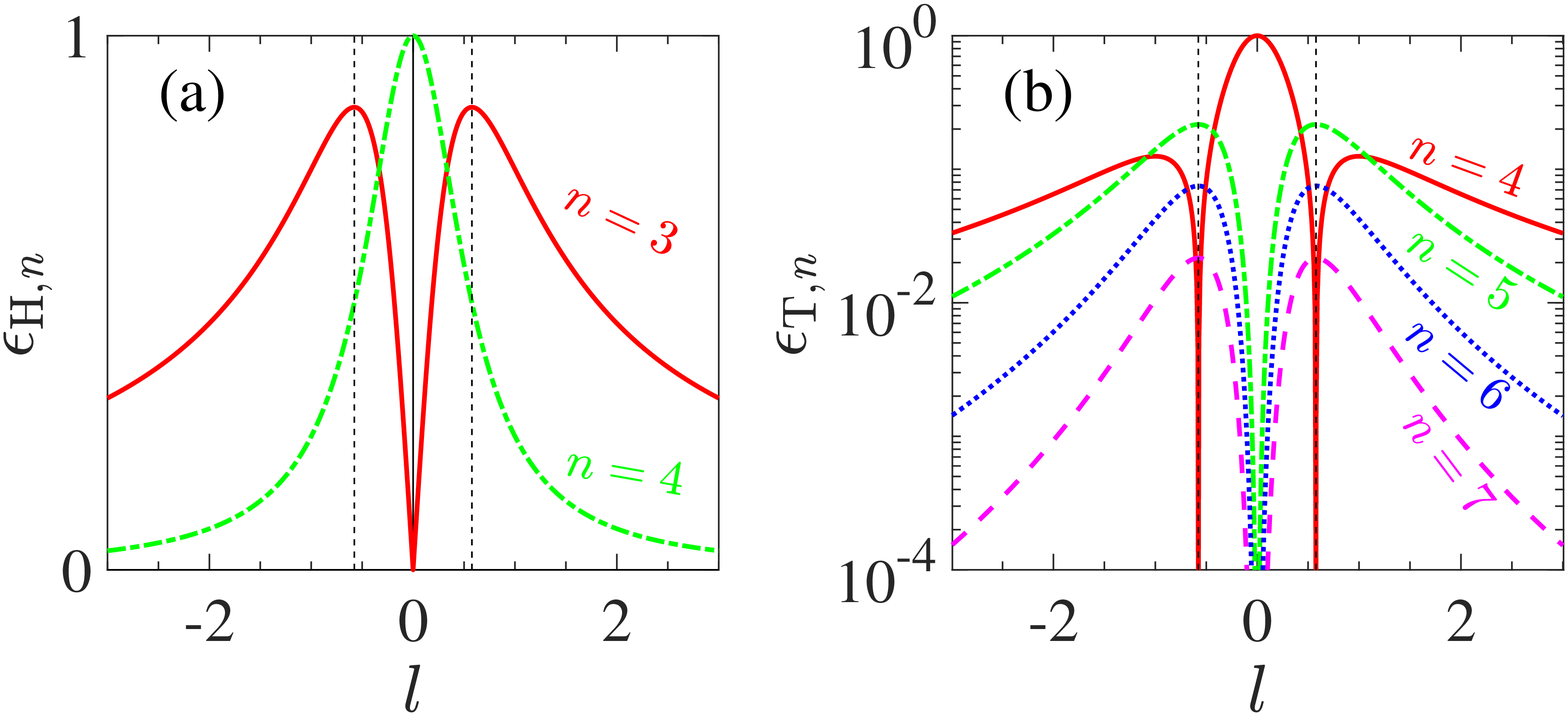}
	\caption{The perturbation strength $\epsilon_{\text{H},n}$ [see Eq.~(\ref{eqPH})] (a) and $\epsilon_{\text{T},n}$ [see Eq.~(\ref{eqPT})] (b) as functions of strain $l$ for different order $n$. Set $\varepsilon=1$ for demo. The vertical dashed lines in both panels are for $l=\pm\sqrt{3}/3$, which are plotted for reference.}\label{figLmabda}
\end{figure}

Figures~\ref{figLmabda}(a) and \ref{figLmabda}(b) give, respectively, the perturbation strength $\epsilon_{\text{H},n}$ and $\epsilon_{\text{T},n}$ as functions of strain $l$, which show a qualitative difference derived from the different choices of $H_0$. For example, at $l=\pm\sqrt{3}/3$, $\epsilon_{\text{H},3}$ reaches the maximum while $\epsilon_{\text{T},4}$ arrives at the minimum value zero, and the order of leading perturbation is also different: the former is the third-order, the latter is the fifth-order. Under this specific strain, if the PHD dominates the system's thermalization behavior, $T_{\text{eq}}$ will be expected to reach a minimum value; otherwise, if the PTD is dominant, $T_{\text{eq}}$ will take a maximum value. Thus, the thermalization behavior of the system near this point can be used to identify which is the most reasonable choice for $H_0$. Besides, from Fig.~\ref{figLmabda}, it is seen that $\epsilon_{\text{H},n}$ and $\epsilon_{\text{T},n}$ are both even functions of $l$, which means that the systems for $\pm l$ have the same dynamic behavior. So we consider mainly $l\geq0$ below. Additionally, it should be pointed out that pressure can adjust the integrability strength of a nonintegrable system, but it does not destroy the integrability of integrable systems like the harmonic chain and the Toda chain \cite{note1}.

\textit{Physical quantities and numerical method}.---For the fixed boundary conditions; i.e., $x_0=p_0=p_{N+1}=0$ and $x_{N+1}=(N+1)b$, the normal modes are defined as
\begin{equation}
    Q_k=\sum_{i=1}^N\sqrt{m_i}q_iU_{i,k},\quad
    P_k=\sum_{i=1}^N\sqrt{m_i}p_iU_{i,k},
\end{equation}
where $q_i=x_i-ib$ is the displacement from the equilibrium
position of the $i$th particle, and $U_{i,k}$ is the $k$th eigenvector. To each mode $k$ one can associate a harmonic energy
\begin{equation}
    E_k=\frac{1}{2}\left(P_k^2+\omega_k^2Q_k^2\right),\quad k=1,\cdots,N,
\end{equation}
where $\omega_k$ is the eigenfrequency of the $k$th normal mode, and a phase $\varphi_k$ defined via
\begin{equation}
   Q_k=\sqrt{2E_k/\omega_k^2}\sin{\left(\varphi_k\right)},~~P_k=\sqrt{2E_k}\cos{\left(\varphi_k\right)}.
\end{equation}
For the monatomic chain, $U_{i,k}$ and $\omega_k$ can be explicitly expressed as
\begin{equation}
  U_{i,k}=\sqrt{\frac{2}{N+1}}\sin\left(\frac{ik\pi}{N+1}\right),~~\omega_k=2\sin\left[\frac{k\pi}{2(N+1)}\right].
\end{equation}
Whereas, for a mass-disordered chain, it is hard to give explicit expressions of $U_{i,k}$ and $\omega_k$. Hence, $U_{i,k}$ and $\omega_k$ are obtained by numerically diagonalizing the harmonic matrix~\cite{1970Localization}. Additionally, the disorder destroys the translation invariance of the system, so the definition of the wave vector is meaningless. Consequently, for the mass-disordered chain, index $k$ only represents the $k$th mode, and it follows an ascending order so that $\omega_k\leq\omega_{k+1}$.

Following the definition of equipartition, one expects
\begin{equation}
  \lim_{T\rightarrow\infty}\bar{E}_k(T)\simeq\varepsilon, \quad k=1,~\cdots,~N,
\end{equation}
where $\bar{E}_k(T)$ denotes the time average of $E_k$ up to time $T$; i.e.,
\begin{equation}\label{eq:EkT}
  \bar{E}_k(T)=\frac{1}{(1-\mu)T}\int_{\mu T}^TE_k(P(t),Q(t))dt.
\end{equation}
Here $\mu\in[0,1)$ controls the size of the time average window, and $\mu=2/3$ is kept fixed throughout the paper.

Based on the defined $\bar{E}_k(T)$, we need to introduce the normalized effective relative number of degrees of freedom~\cite{PhysRevA.31.1039},
\begin{equation}\label{eq:xi}
  \xi(t)=N^{-1}e^{\eta(t)},
\end{equation}
to measure how close the system is to equipartition, where $\eta(t)=-\sum_{k=1}^{N}w_k(t)\log[w_k(t)]$ is the spectral entropy and $  w_k(t)=\bar{E}_k(t)/\left[\sum_{j=1}^N\bar{E}_j(t)\right]$.
When the thermalized state is approached, $\xi$ will saturate at $1$.

In our simulations, the motion equations are integrated by the eighth-order Yoshida algorithm~\cite{YOSHIDA1990262}.~The time step is $\Delta t=0.05$; the corresponding relative error in energy conservation is less than $10^{-5}$. To suppress fluctuations, the average is done over $24$ phases uniformly distributed in $[0,2\pi]$, and hereafter we use $\langle\cdot\rangle$ to denote the ensemble average results. The lattice constant $a=10$ and the mass-disordered strength $\Delta m=0.2$ are kept fixed. Besides, the lowest $10\%$ of frequency modes are excited initially throughout all calculations. We have checked and verified that no qualitative difference will be resulted in when the percentage of the excited modes is changed.

\begin{figure}[t]
  \centering
  \includegraphics[width=1\columnwidth]{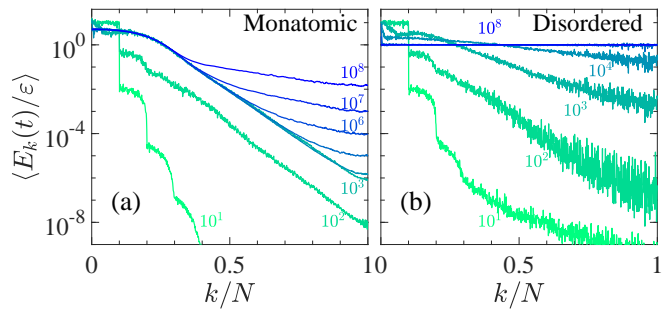}\\
  \caption{The function~$\langle E_k(t)/\varepsilon\rangle$ versus $k/N$ at different times for the monatomic (a) and the mass-disordered (b) FPUT-$\beta$ chains with strain $l=\sqrt{3}/3$, in semi-log scale. The system size~$N=1023$, and the energy density~$\varepsilon=0.001$ are kept fixed with ensemble average measurements for $24$ different random choices of the phases.}\label{figEkT}
\end{figure}

\textit{Numerical results}.---In Figs.~\ref{figEkT}(a) and \ref{figEkT}(c), we show, respectively, the energy $\langle E_k(t)/\varepsilon\rangle$ against $k/N$ at selected times for the monatomic chains and the mass-disordered chains of FPUT-$\beta$ with strain $l=\sqrt{3}/3$. Note that the energy assigned to the low-frequency modes at the initial time gradually transfers to the high-frequency modes. It can also be seen that the mass-disordered chain tends to thermalized state faster than the monatomic chain.

\begin{figure}[t]
  \centering
  \includegraphics[width=1\columnwidth]{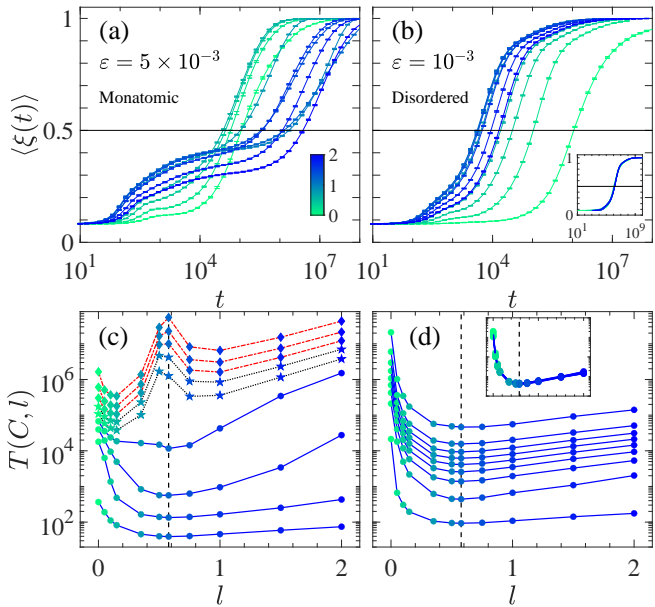}\\
  \caption{The function of $\langle\xi(t)\rangle$ for the monatomic (a) and the mass-disordered (b) FPUT-$\beta$ chains with different strains, $l$, in semi-log scale. The colorbar in panel (a) is for $l$, which is the applied for all panels. The time $T(C,l)$ as a function of strain $l$ for the monatomic (c) and the mass-disordered (d) FPUT-$\beta$ chains, which are measured directly from the data in panels (a) and (b), respectively. The curves from bottom to top are for $C=0.1$, $0.2$, $\cdots$, $0.9$, respectively. The vertical dashed line is for $l=\sqrt{3}/3$, which is plotted for reference. In panel (c), inset: Same as the main panel but the curves are shifted properly in the horizontal direction (with that for $l=0$ unshifted) so that they perfectly overlap with each other. In panel (d), inset: Same as the main panel but the curves are shifted properly in the vertical direction (with that for $C=0.9$ unshifted). These are ensemble average measurements of $24$ different random choices of phases.~The system size~$N=2047$. The strength of disorder $\Delta m=0.2$ is fixed for the mass-disordered case.}\label{figXiTeq}
\end{figure}

To observe more detail of thermalization dynamics, we study the time evolution of $\langle \xi(t)\rangle$ defined by Eq.~(\ref{eq:xi}). In Figs.~\ref{figXiTeq}(a) and \ref{figXiTeq}(b), we show the results for the monatomic chains and the mass-disordered chains, respectively. At first glance, on a sufficiently large timescale all values of $\langle \xi(t)\rangle$ tend to $1$, which implies that the energy equipartition will be achieved finally. Meanwhile, it is also seen that there are qualitative differences. All curves for the mass-disordered chains [see Fig.~\ref{figXiTeq}(b)] are approximately the same with very similar sigmoidal profiles, and they can overlap with each other upon suitable shifts (see the inset in main panel), which suggests that the thermalization dynamics of the whole process is consistent under different strains. But the curves for the monatomic chains [see Fig.~\ref{figXiTeq}(a)] are impossible to overlap completely by a suitable shift, so the thermalization dynamics is strain-dependent and time-dependent.

To clearly show the dependence of the thermalization process on strain, we define the time $T(C,l)$ as a function of $C$ and $l$, where $T$ is the time when $\langle \xi(t)\rangle$ reaches the threshold value $C$, i.e., $\langle \xi(T)\rangle=C$, for a given $l$. Figure~\ref{figXiTeq}(c) presents the results for the monatomic chains. Notice that the curves of $T(C,l)$ versus $l$ change qualitatively as $C$ increases, such as the curves for $C\leq 0.4$ change in exactly the same way, and reach the minimum value at $l=\sqrt{3}/3$, which fits the prediction when the PHD dominates the thermalization process; the curves for $C\geq0.7$ are qualitatively consistent, and take the maximum value at $l=\sqrt{3}/3$, which agrees with the expectations when the PTD is dominant for the thermalization. These results suggest that thermalization of the monatomic chain is dominated by the PTD, although it shows the behavior of the PHD in a short time. As a contrast, we show the results for the mass-disordered chains in Fig.~\ref{figXiTeq}(d). From the main panel and the inset, it can be seen that all the curves for $C=0.2$, $\cdots$, $0.9$ have almost exactly the same shape, and all curves reach the minimum at $l=\sqrt{3}/3$, which implies that the PHD dominates the whole process of thermalization. This result is consistent with the fact that the mass-disorder destroys the integrability of the Toda, the harmonic system with mass-disorder is the reference integrable system of the mass-disordered chain. Thus, the thermalization behavior of the mass-disordered chain is always determined by the PHD, which results in good scaling behavior of the curves in Figs.~\ref{figXiTeq}(b) and \ref{figXiTeq}(d). Next, we will explore the dependence of thermalization time $T_\text{eq}$ on strain $l$.

\begin{figure}[t]
  \centering
  \includegraphics[width=1\columnwidth]{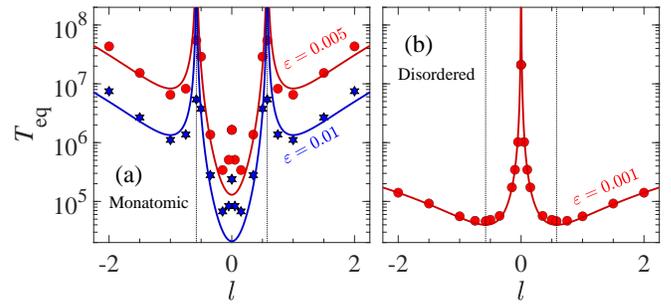}\\
  \caption{(a) The thermalization time $T_\text{eq}$ as a function of strain $l$ for the stretched monatomic FPUT-$\beta$ chain with different energy densities in semi-log scale. The solid lines with the function curve of form $\sim\epsilon_{\text{T},4}^{-2}$ are drawn for reference. (b) The results for the mass-disordered FPUT-$\beta$ chain under strain. The solid line with the function curve of form $\sim\epsilon_{\text{H},3}^{-2}$ is drawn for reference. The black vertical dashed lines in panels (a) and (b) are for $l=\pm\sqrt{3}/3$, which are plotted for reference.}\label{figTeq}
\end{figure}

Strictly speaking, $T_\text{eq}$ is defined as that when $\langle \xi(t)\rangle$ reaches $1$. However, we usually only care about the scaling behavior of $T_\text{eq}$, but not the specific value. Therefore, if $\langle \xi(t)\rangle$ has a good scaling behavior near $1$, $T_\text{eq}$ is generally defined as the time when $\langle \xi(t)\rangle$ reaches a certain threshold less than $1$ in order to save the cost of calculation~\cite{2008LNP728G}. Here, we define $T_\text{eq}$ as $\langle \xi(t)\rangle=0.9$ for the two kinds of chains, i.e., $T_\text{eq}(l)=T(0.9,l)$. The scalability of the curves for $C\geq0.7$ in Figs.~\ref{figXiTeq}(c) and \ref{figXiTeq}(d) state that the definition does not affect the dependence of $T_\text{eq}$ on $l$ here. Figure~\ref{figTeq}(a) shows the results for the stretched monatomic FPUT-$\beta$ chains with different energy densities. The solid lines with the function curve of the form $\sim\epsilon_{\text{T},4}^{-2}$ are plotted for reference. Note that two remarkable peaks locate at $l=\pm\sqrt{3}/3$, which are predictable by the fact that $\epsilon_{\text{T},4}(\pm\sqrt{3}/3)=0$ [see Fig.~\ref{figLmabda}(b)], i.e., the system is of the fifth-order perturbation at these points, so the integrability is enhanced, which makes the system more difficult to be thermalized. Overall, the data points and the reference lines fit pretty well except for some points near $l=0$. When $l$ approaches $0$, the inter-particle interaction potential of the system tends to a symmetric function, which has a strong finite-size effect, thus it needs a larger size to tend to the theoretical prediction. The phenomenon that the symmetric model has a stronger finite-size effect than the asymmetric model has been reported in the literatures~\cite{Benettin2011,Our2018,PhysRevE.100.052102}. Figure~\ref{figTeq}(b) presents the results for the mass-disordered chains. It can be seen that all numerical points are well covered by the solid line with the function curve of the form $\sim\epsilon_{\text{H},3}^{-2}$, which strongly suggests that the thermalization process of the stretched mass-disordered FPUT-$\beta$ chain is completely ruled by the third-order anharmonicity, i.e., the PHD. The result that thermalization time is inversely proportional to the strength of leading perturbation is consistent with the thermalization law recently found in other nonlinear systems~\cite{Our2018,Fu_2019,PhysRevE.100.052102,PhysRevLett.124.186401,PhysRevE.94.062104,Pistone2018,2020arXiv200503478W}.

\textit{Conclusions and discussions}.---In summary, we have studied the thermalization properties of the FPUT-$\beta$ chains under strain in both monatomic and mass-disordered cases, in the thermodynamic limit. Our results show that as long as one selects a suitable reference integrable system to accurately define the perturbation strength, $\epsilon$, which guarantees that the ability of the system to be thermalized is depicted faithfully, the thermalization behavior of the system under stress follows the law: $T_\text{eq}\propto\epsilon^{-2}$.

By comparing the thermalization processes of the monatomic chains and the mass-disordered chains (see again Fig.~\ref{figXiTeq}), it is found that for the monatomic chain, the PHD is responsible for the thermalization in a short time, but the PTD governs the thermalization behavior finally, while the PHD rules the whole process of thermalization of the mass-disordered chain. All these results strongly indicate that pressure can dramatically change the integrability of the system, which plays a crucial role in the thermal transport properties~\cite{PhysRevE.90.032134,lepri2016thermal,PhysRevE.97.010103,Livi_2020,PhysRevLett.125.040604}. Hence, our results provide a new insight into understanding why and how pressure changes the thermal transport properties of the system.

\begin{acknowledgments}
We acknowledge support by the NSFC (Grant Nos.~12005156, 11975190, 11975189, 12047501, 12064037, 11665019, 11964031, and 11764035), and by the Natural Science Foundation of Gansu Province (Grant Nos.~20JR5RA494, and 21JR1RE289), and by the Innovation Fund for Colleges and Universities from Department of Education of Gansu Province, China (Grant No.~2020B-169). The work was carried out at the National Supercomputer Center in Tianjin, and the calculations were performed on TianHe-1(A).
\end{acknowledgments}

\bibliography{Reference}

\begin{thebibliography}{49}%
\makeatletter
\providecommand \@ifxundefined [1]{%
 \@ifx{#1\undefined}
}%
\providecommand \@ifnum [1]{%
 \ifnum #1\expandafter \@firstoftwo
 \else \expandafter \@secondoftwo
 \fi
}%
\providecommand \@ifx [1]{%
 \ifx #1\expandafter \@firstoftwo
 \else \expandafter \@secondoftwo
 \fi
}%
\providecommand \natexlab [1]{#1}%
\providecommand \enquote  [1]{``#1''}%
\providecommand \bibnamefont  [1]{#1}%
\providecommand \bibfnamefont [1]{#1}%
\providecommand \citenamefont [1]{#1}%
\providecommand \href@noop [0]{\@secondoftwo}%
\providecommand \href [0]{\begingroup \@sanitize@url \@href}%
\providecommand \@href[1]{\@@startlink{#1}\@@href}%
\providecommand \@@href[1]{\endgroup#1\@@endlink}%
\providecommand \@sanitize@url [0]{\catcode `\\12\catcode `\$12\catcode
  `\&12\catcode `\#12\catcode `\^12\catcode `\_12\catcode `\%12\relax}%
\providecommand \@@startlink[1]{}%
\providecommand \@@endlink[0]{}%
\providecommand \url  [0]{\begingroup\@sanitize@url \@url }%
\providecommand \@url [1]{\endgroup\@href {#1}{\urlprefix }}%
\providecommand \urlprefix  [0]{URL }%
\providecommand \Eprint [0]{\href }%
\providecommand \doibase [0]{http://dx.doi.org/}%
\providecommand \selectlanguage [0]{\@gobble}%
\providecommand \bibinfo  [0]{\@secondoftwo}%
\providecommand \bibfield  [0]{\@secondoftwo}%
\providecommand \translation [1]{[#1]}%
\providecommand \BibitemOpen [0]{}%
\providecommand \bibitemStop [0]{}%
\providecommand \bibitemNoStop [0]{.\EOS\space}%
\providecommand \EOS [0]{\spacefactor3000\relax}%
\providecommand \BibitemShut  [1]{\csname bibitem#1\endcsname}%
\let\auto@bib@innerbib\@empty
\bibitem [{\citenamefont {Drozdov}\ \emph {et~al.}(2015)\citenamefont
  {Drozdov}, \citenamefont {Eremets}, \citenamefont {Troyan}, \citenamefont
  {Ksenofontov},\ and\ \citenamefont {Shylin}}]{RN334}%
  \BibitemOpen
  \bibfield  {author} {\bibinfo {author} {\bibfnamefont {A.~P.}\ \bibnamefont
  {Drozdov}}, \bibinfo {author} {\bibfnamefont {M.~I.}\ \bibnamefont
  {Eremets}}, \bibinfo {author} {\bibfnamefont {I.~A.}\ \bibnamefont {Troyan}},
  \bibinfo {author} {\bibfnamefont {V.}~\bibnamefont {Ksenofontov}}, \ and\
  \bibinfo {author} {\bibfnamefont {S.~I.}\ \bibnamefont {Shylin}},\ }\href
  {\doibase 10.1038/nature14964} {\bibfield  {journal} {\bibinfo  {journal}
  {Nature}\ }\textbf {\bibinfo {volume} {525}},\ \bibinfo {pages} {73}
  (\bibinfo {year} {2015})}\BibitemShut {NoStop}%
\bibitem [{\citenamefont {Drozdov}\ \emph {et~al.}(2019)\citenamefont
  {Drozdov}, \citenamefont {Kong}, \citenamefont {Minkov}, \citenamefont
  {Besedin}, \citenamefont {Kuzovnikov}, \citenamefont {Mozaffari},
  \citenamefont {Balicas}, \citenamefont {Balakirev}, \citenamefont {Graf},
  \citenamefont {Prakapenka}, \citenamefont {Greenberg}, \citenamefont
  {Knyazev}, \citenamefont {Tkacz},\ and\ \citenamefont {Eremets}}]{RN335}%
  \BibitemOpen
  \bibfield  {author} {\bibinfo {author} {\bibfnamefont {A.~P.}\ \bibnamefont
  {Drozdov}}, \bibinfo {author} {\bibfnamefont {P.~P.}\ \bibnamefont {Kong}},
  \bibinfo {author} {\bibfnamefont {V.~S.}\ \bibnamefont {Minkov}}, \bibinfo
  {author} {\bibfnamefont {S.~P.}\ \bibnamefont {Besedin}}, \bibinfo {author}
  {\bibfnamefont {M.~A.}\ \bibnamefont {Kuzovnikov}}, \bibinfo {author}
  {\bibfnamefont {S.}~\bibnamefont {Mozaffari}}, \bibinfo {author}
  {\bibfnamefont {L.}~\bibnamefont {Balicas}}, \bibinfo {author} {\bibfnamefont
  {F.~F.}\ \bibnamefont {Balakirev}}, \bibinfo {author} {\bibfnamefont {D.~E.}\
  \bibnamefont {Graf}}, \bibinfo {author} {\bibfnamefont {V.~B.}\ \bibnamefont
  {Prakapenka}}, \bibinfo {author} {\bibfnamefont {E.}~\bibnamefont
  {Greenberg}}, \bibinfo {author} {\bibfnamefont {D.~A.}\ \bibnamefont
  {Knyazev}}, \bibinfo {author} {\bibfnamefont {M.}~\bibnamefont {Tkacz}}, \
  and\ \bibinfo {author} {\bibfnamefont {M.~I.}\ \bibnamefont {Eremets}},\
  }\href {\doibase 10.1038/s41586-019-1201-8} {\bibfield  {journal} {\bibinfo
  {journal} {Nature}\ }\textbf {\bibinfo {volume} {569}},\ \bibinfo {pages}
  {528} (\bibinfo {year} {2019})}\BibitemShut {NoStop}%
\bibitem [{\citenamefont {Liu}\ \emph {et~al.}(2019)\citenamefont {Liu},
  \citenamefont {Wang}, \citenamefont {Yi}, \citenamefont {Kim}, \citenamefont
  {Kim},\ and\ \citenamefont {Cho}}]{PhysRevB.99.140501}%
  \BibitemOpen
  \bibfield  {author} {\bibinfo {author} {\bibfnamefont {L.}~\bibnamefont
  {Liu}}, \bibinfo {author} {\bibfnamefont {C.}~\bibnamefont {Wang}}, \bibinfo
  {author} {\bibfnamefont {S.}~\bibnamefont {Yi}}, \bibinfo {author}
  {\bibfnamefont {K.~W.}\ \bibnamefont {Kim}}, \bibinfo {author} {\bibfnamefont
  {J.}~\bibnamefont {Kim}}, \ and\ \bibinfo {author} {\bibfnamefont {J.-H.}\
  \bibnamefont {Cho}},\ }\href {\doibase 10.1103/PhysRevB.99.140501} {\bibfield
   {journal} {\bibinfo  {journal} {Phys. Rev. B}\ }\textbf {\bibinfo {volume}
  {99}},\ \bibinfo {pages} {140501(R)} (\bibinfo {year} {2019})}\BibitemShut
  {NoStop}%
\bibitem [{\citenamefont {Errea}\ \emph {et~al.}(2020)\citenamefont {Errea},
  \citenamefont {Belli}, \citenamefont {Monacelli}, \citenamefont {Sanna},
  \citenamefont {Koretsune}, \citenamefont {Tadano}, \citenamefont {Bianco},
  \citenamefont {Calandra}, \citenamefont {Arita}, \citenamefont {Mauri},\ and\
  \citenamefont {Flores-Livas}}]{RN336}%
  \BibitemOpen
  \bibfield  {author} {\bibinfo {author} {\bibfnamefont {I.}~\bibnamefont
  {Errea}}, \bibinfo {author} {\bibfnamefont {F.}~\bibnamefont {Belli}},
  \bibinfo {author} {\bibfnamefont {L.}~\bibnamefont {Monacelli}}, \bibinfo
  {author} {\bibfnamefont {A.}~\bibnamefont {Sanna}}, \bibinfo {author}
  {\bibfnamefont {T.}~\bibnamefont {Koretsune}}, \bibinfo {author}
  {\bibfnamefont {T.}~\bibnamefont {Tadano}}, \bibinfo {author} {\bibfnamefont
  {R.}~\bibnamefont {Bianco}}, \bibinfo {author} {\bibfnamefont
  {M.}~\bibnamefont {Calandra}}, \bibinfo {author} {\bibfnamefont
  {R.}~\bibnamefont {Arita}}, \bibinfo {author} {\bibfnamefont
  {F.}~\bibnamefont {Mauri}}, \ and\ \bibinfo {author} {\bibfnamefont {J.~A.}\
  \bibnamefont {Flores-Livas}},\ }\href {\doibase 10.1038/s41586-020-1955-z}
  {\bibfield  {journal} {\bibinfo  {journal} {Nature}\ }\textbf {\bibinfo
  {volume} {578}},\ \bibinfo {pages} {66} (\bibinfo {year} {2020})}\BibitemShut
  {NoStop}%
\bibitem [{\citenamefont {Flores-Livas}\ \emph {et~al.}(2020)\citenamefont
  {Flores-Livas}, \citenamefont {Boeri}, \citenamefont {Sanna}, \citenamefont
  {Profeta}, \citenamefont {Arita},\ and\ \citenamefont
  {Eremets}}]{FLORESLIVAS20201}%
  \BibitemOpen
  \bibfield  {author} {\bibinfo {author} {\bibfnamefont {J.~A.}\ \bibnamefont
  {Flores-Livas}}, \bibinfo {author} {\bibfnamefont {L.}~\bibnamefont {Boeri}},
  \bibinfo {author} {\bibfnamefont {A.}~\bibnamefont {Sanna}}, \bibinfo
  {author} {\bibfnamefont {G.}~\bibnamefont {Profeta}}, \bibinfo {author}
  {\bibfnamefont {R.}~\bibnamefont {Arita}}, \ and\ \bibinfo {author}
  {\bibfnamefont {M.}~\bibnamefont {Eremets}},\ }\href {\doibase
  https://doi.org/10.1016/j.physrep.2020.02.003} {\bibfield  {journal}
  {\bibinfo  {journal} {Phys. Rep.}\ }\textbf {\bibinfo {volume} {856}},\
  \bibinfo {pages} {1} (\bibinfo {year} {2020})}\BibitemShut {NoStop}%
\bibitem [{\citenamefont {Pickard}\ \emph {et~al.}(2020)\citenamefont
  {Pickard}, \citenamefont {Errea},\ and\ \citenamefont
  {Eremets}}]{Pickard2020}%
  \BibitemOpen
  \bibfield  {author} {\bibinfo {author} {\bibfnamefont {C.~J.}\ \bibnamefont
  {Pickard}}, \bibinfo {author} {\bibfnamefont {I.}~\bibnamefont {Errea}}, \
  and\ \bibinfo {author} {\bibfnamefont {M.~I.}\ \bibnamefont {Eremets}},\
  }\href {\doibase 10.1146/annurev-conmatphys-031218-013413} {\bibfield
  {journal} {\bibinfo  {journal} {Annu. Rev. Condens. Matter Phys.}\ }\textbf
  {\bibinfo {volume} {11}},\ \bibinfo {pages} {57} (\bibinfo {year}
  {2020})}\BibitemShut {NoStop}%
\bibitem [{\citenamefont {Ravichandran}\ and\ \citenamefont
  {Broido}(2019)}]{RN337}%
  \BibitemOpen
  \bibfield  {author} {\bibinfo {author} {\bibfnamefont {N.~K.}\ \bibnamefont
  {Ravichandran}}\ and\ \bibinfo {author} {\bibfnamefont {D.}~\bibnamefont
  {Broido}},\ }\href {\doibase 10.1038/s41467-019-08713-0} {\bibfield
  {journal} {\bibinfo  {journal} {Nat. Commun.}\ }\textbf {\bibinfo {volume}
  {10}},\ \bibinfo {pages} {827} (\bibinfo {year} {2019})}\BibitemShut
  {NoStop}%
\bibitem [{\citenamefont {Mohiuddin}\ \emph {et~al.}(2009)\citenamefont
  {Mohiuddin}, \citenamefont {Lombardo}, \citenamefont {Nair}, \citenamefont
  {Bonetti}, \citenamefont {Savini}, \citenamefont {Jalil}, \citenamefont
  {Bonini}, \citenamefont {Basko}, \citenamefont {Galiotis}, \citenamefont
  {Marzari}, \citenamefont {Novoselov}, \citenamefont {Geim},\ and\
  \citenamefont {Ferrari}}]{PhysRevB.79.205433}%
  \BibitemOpen
  \bibfield  {author} {\bibinfo {author} {\bibfnamefont {T.~M.~G.}\
  \bibnamefont {Mohiuddin}}, \bibinfo {author} {\bibfnamefont {A.}~\bibnamefont
  {Lombardo}}, \bibinfo {author} {\bibfnamefont {R.~R.}\ \bibnamefont {Nair}},
  \bibinfo {author} {\bibfnamefont {A.}~\bibnamefont {Bonetti}}, \bibinfo
  {author} {\bibfnamefont {G.}~\bibnamefont {Savini}}, \bibinfo {author}
  {\bibfnamefont {R.}~\bibnamefont {Jalil}}, \bibinfo {author} {\bibfnamefont
  {N.}~\bibnamefont {Bonini}}, \bibinfo {author} {\bibfnamefont {D.~M.}\
  \bibnamefont {Basko}}, \bibinfo {author} {\bibfnamefont {C.}~\bibnamefont
  {Galiotis}}, \bibinfo {author} {\bibfnamefont {N.}~\bibnamefont {Marzari}},
  \bibinfo {author} {\bibfnamefont {K.~S.}\ \bibnamefont {Novoselov}}, \bibinfo
  {author} {\bibfnamefont {A.~K.}\ \bibnamefont {Geim}}, \ and\ \bibinfo
  {author} {\bibfnamefont {A.~C.}\ \bibnamefont {Ferrari}},\ }\href {\doibase
  10.1103/PhysRevB.79.205433} {\bibfield  {journal} {\bibinfo  {journal} {Phys.
  Rev. B}\ }\textbf {\bibinfo {volume} {79}},\ \bibinfo {pages} {205433}
  (\bibinfo {year} {2009})}\BibitemShut {NoStop}%
\bibitem [{\citenamefont {Li}\ \emph {et~al.}(2010)\citenamefont {Li},
  \citenamefont {Maute}, \citenamefont {Dunn},\ and\ \citenamefont
  {Yang}}]{PhysRevB.81.245318}%
  \BibitemOpen
  \bibfield  {author} {\bibinfo {author} {\bibfnamefont {X.}~\bibnamefont
  {Li}}, \bibinfo {author} {\bibfnamefont {K.}~\bibnamefont {Maute}}, \bibinfo
  {author} {\bibfnamefont {M.~L.}\ \bibnamefont {Dunn}}, \ and\ \bibinfo
  {author} {\bibfnamefont {R.}~\bibnamefont {Yang}},\ }\href {\doibase
  10.1103/PhysRevB.81.245318} {\bibfield  {journal} {\bibinfo  {journal} {Phys.
  Rev. B}\ }\textbf {\bibinfo {volume} {81}},\ \bibinfo {pages} {245318}
  (\bibinfo {year} {2010})}\BibitemShut {NoStop}%
\bibitem [{\citenamefont {Yuan}\ \emph {et~al.}(2019)\citenamefont {Yuan},
  \citenamefont {Zhang}, \citenamefont {Li},\ and\ \citenamefont
  {Tang}}]{C8CP06414H}%
  \BibitemOpen
  \bibfield  {author} {\bibinfo {author} {\bibfnamefont {K.}~\bibnamefont
  {Yuan}}, \bibinfo {author} {\bibfnamefont {X.}~\bibnamefont {Zhang}},
  \bibinfo {author} {\bibfnamefont {L.}~\bibnamefont {Li}}, \ and\ \bibinfo
  {author} {\bibfnamefont {D.}~\bibnamefont {Tang}},\ }\href {\doibase
  10.1039/C8CP06414H} {\bibfield  {journal} {\bibinfo  {journal} {Phys. Chem.
  Chem. Phys.}\ }\textbf {\bibinfo {volume} {21}},\ \bibinfo {pages} {468}
  (\bibinfo {year} {2019})}\BibitemShut {NoStop}%
\bibitem [{\citenamefont {Jiang}\ \emph {et~al.}(2017)\citenamefont {Jiang},
  \citenamefont {Fu}, \citenamefont {Chen},\ and\ \citenamefont
  {Zhao}}]{RN338}%
  \BibitemOpen
  \bibfield  {author} {\bibinfo {author} {\bibfnamefont {J.}~\bibnamefont
  {Jiang}}, \bibinfo {author} {\bibfnamefont {W.}~\bibnamefont {Fu}}, \bibinfo
  {author} {\bibfnamefont {J.}~\bibnamefont {Chen}}, \ and\ \bibinfo {author}
  {\bibfnamefont {H.}~\bibnamefont {Zhao}},\ }\href {\doibase
  10.1007/s11433-017-9041-8} {\bibfield  {journal} {\bibinfo  {journal} {Sci.
  China-Phys. Mech. Astron.}\ }\textbf {\bibinfo {volume} {60}},\ \bibinfo
  {pages} {070512} (\bibinfo {year} {2017})}\BibitemShut {NoStop}%
\bibitem [{\citenamefont {S\"a\"askilahti}\ \emph {et~al.}(2015)\citenamefont
  {S\"a\"askilahti}, \citenamefont {Oksanen}, \citenamefont {Volz},\ and\
  \citenamefont {Tulkki}}]{PhysRevB.92.245411}%
  \BibitemOpen
  \bibfield  {author} {\bibinfo {author} {\bibfnamefont {K.}~\bibnamefont
  {S\"a\"askilahti}}, \bibinfo {author} {\bibfnamefont {J.}~\bibnamefont
  {Oksanen}}, \bibinfo {author} {\bibfnamefont {S.}~\bibnamefont {Volz}}, \
  and\ \bibinfo {author} {\bibfnamefont {J.}~\bibnamefont {Tulkki}},\ }\href
  {\doibase 10.1103/PhysRevB.92.245411} {\bibfield  {journal} {\bibinfo
  {journal} {Phys. Rev. B}\ }\textbf {\bibinfo {volume} {92}},\ \bibinfo
  {pages} {245411} (\bibinfo {year} {2015})}\BibitemShut {NoStop}%
\bibitem [{\citenamefont {Jiang}\ and\ \citenamefont
  {Zhao}(2016)}]{Jiang_2016}%
  \BibitemOpen
  \bibfield  {author} {\bibinfo {author} {\bibfnamefont {J.}~\bibnamefont
  {Jiang}}\ and\ \bibinfo {author} {\bibfnamefont {H.}~\bibnamefont {Zhao}},\
  }\href {\doibase 10.1088/1742-5468/2016/09/093208} {\bibfield  {journal}
  {\bibinfo  {journal} {J. Stat. Mech. Theory Exp.}\ }\textbf {\bibinfo
  {volume} {2016}},\ \bibinfo {pages} {093208} (\bibinfo {year}
  {2016})}\BibitemShut {NoStop}%
\bibitem [{\citenamefont {Sato}(2016)}]{PhysRevE.94.012115}%
  \BibitemOpen
  \bibfield  {author} {\bibinfo {author} {\bibfnamefont {D.~SK.}\ \bibnamefont
  {Sato}},\ }\href {\doibase 10.1103/PhysRevE.94.012115} {\bibfield  {journal}
  {\bibinfo  {journal} {Phys. Rev. E}\ }\textbf {\bibinfo {volume} {94}},\
  \bibinfo {pages} {012115} (\bibinfo {year} {2016})}\BibitemShut {NoStop}%
\bibitem [{\citenamefont {Sato}(2020)}]{PhysRevE.102.012111}%
  \BibitemOpen
  \bibfield  {author} {\bibinfo {author} {\bibfnamefont {D.~SK.}\ \bibnamefont
  {Sato}},\ }\href {\doibase 10.1103/PhysRevE.102.012111} {\bibfield  {journal}
  {\bibinfo  {journal} {Phys. Rev. E}\ }\textbf {\bibinfo {volume} {102}},\
  \bibinfo {pages} {012111} (\bibinfo {year} {2020})}\BibitemShut {NoStop}%
\bibitem [{\citenamefont {Khinchin}(1949)}]{1949Khinchin}%
  \BibitemOpen
  \bibfield  {author} {\bibinfo {author} {\bibfnamefont {A.~I.}\ \bibnamefont
  {Khinchin}},\ }\href@noop {} {\emph {\bibinfo {title} {Mathematical
  Foundations of Statistical Mechanics}}}\ (\bibinfo  {publisher} {Dover, New
  York},\ \bibinfo {year} {1949})\BibitemShut {NoStop}%
\bibitem [{\citenamefont {Fermi}\ \emph {et~al.}(1955)\citenamefont {Fermi},
  \citenamefont {Pasta},\ and\ \citenamefont {Ulam}}]{Fermi1955}%
  \BibitemOpen
  \bibfield  {author} {\bibinfo {author} {\bibfnamefont {E.}~\bibnamefont
  {Fermi}}, \bibinfo {author} {\bibfnamefont {J.}~\bibnamefont {Pasta}}, \ and\
  \bibinfo {author} {\bibfnamefont {S.}~\bibnamefont {Ulam}},\ }\href@noop {}
  {\bibfield  {journal} {\bibinfo  {journal} {Los Alamos Scientific Laboratory,
  Report No. LA-1940}\ } (\bibinfo {year} {1955})}\BibitemShut {NoStop}%
\bibitem [{\citenamefont {Dauxois}(2008)}]{dauxois:ensl-00202296}%
  \BibitemOpen
  \bibfield  {author} {\bibinfo {author} {\bibfnamefont {T.}~\bibnamefont
  {Dauxois}},\ }\href {\doibase 10.1063/1.2835154} {\bibfield  {journal}
  {\bibinfo  {journal} {{Phys. Today}}\ }\textbf {\bibinfo {volume} {61}},\
  \bibinfo {pages} {55} (\bibinfo {year} {2008})}\BibitemShut {NoStop}%
\bibitem [{Cha(2005)}]{Chaos2005}%
  \BibitemOpen
  \href@noop {} {\bibfield  {journal} {\bibinfo  {journal} {Chaos Focus Issue:
  The ``Fermi-Pasta-Ulam'' problem---the first 50 years,~Chaos}\ }\textbf
  {\bibinfo {volume} {15}} (\bibinfo {year} {2005})}\BibitemShut {NoStop}%
\bibitem [{\citenamefont {{Gallavotti}}(2008)}]{2008LNP728G}%
  \BibitemOpen
  \bibinfo {editor} {\bibfnamefont {G.}~\bibnamefont {{Gallavotti}}},\ ed.,\
  \href@noop {} {\emph {\bibinfo {title} {The Fermi-Pasta-Ulam Problem: A
  Status Report}}},\ \bibinfo {series} {Lect. Notes Phys.}, Vol.\ \bibinfo
  {volume} {728}\ (\bibinfo  {publisher} {Springer, New York},\ \bibinfo {year}
  {2008})\BibitemShut {NoStop}%
\bibitem [{\citenamefont {Zabusky}(2005)}]{doi:10.1063/1.1861554}%
  \BibitemOpen
  \bibfield  {author} {\bibinfo {author} {\bibfnamefont {N.~J.}\ \bibnamefont
  {Zabusky}},\ }\href {\doibase 10.1063/1.1861554} {\bibfield  {journal}
  {\bibinfo  {journal} {Chaos}\ }\textbf {\bibinfo {volume} {15}},\ \bibinfo
  {pages} {015102} (\bibinfo {year} {2005})}\BibitemShut {NoStop}%
\bibitem [{\citenamefont {Campbell}\ \emph {et~al.}(2005)\citenamefont
  {Campbell}, \citenamefont {Rosenau},\ and\ \citenamefont
  {Zaslavsky}}]{doi:10.1063/1.1889345}%
  \BibitemOpen
  \bibfield  {author} {\bibinfo {author} {\bibfnamefont {D.~K.}\ \bibnamefont
  {Campbell}}, \bibinfo {author} {\bibfnamefont {P.}~\bibnamefont {Rosenau}}, \
  and\ \bibinfo {author} {\bibfnamefont {G.~M.}\ \bibnamefont {Zaslavsky}},\
  }\href {\doibase 10.1063/1.1889345} {\bibfield  {journal} {\bibinfo
  {journal} {Chaos}\ }\textbf {\bibinfo {volume} {15}},\ \bibinfo {pages}
  {015101} (\bibinfo {year} {2005})}\BibitemShut {NoStop}%
\bibitem [{\citenamefont {Porter}\ \emph {et~al.}(2009)\citenamefont {Porter},
  \citenamefont {Zabusky}, \citenamefont {Hu},\ and\ \citenamefont
  {Campbell}}]{Porter2009Fermi}%
  \BibitemOpen
  \bibfield  {author} {\bibinfo {author} {\bibfnamefont {M.~A.}\ \bibnamefont
  {Porter}}, \bibinfo {author} {\bibfnamefont {N.~J.}\ \bibnamefont {Zabusky}},
  \bibinfo {author} {\bibfnamefont {B.}~\bibnamefont {Hu}}, \ and\ \bibinfo
  {author} {\bibfnamefont {D.~K.}\ \bibnamefont {Campbell}},\ }\href {\doibase
  10.1511/2009.78.214} {\bibfield  {journal} {\bibinfo  {journal} {American
  Scientist}\ }\textbf {\bibinfo {volume} {97}},\ \bibinfo {pages} {214}
  (\bibinfo {year} {2009})}\BibitemShut {NoStop}%
\bibitem [{\citenamefont {Onorato}\ \emph {et~al.}(2015)\citenamefont
  {Onorato}, \citenamefont {Vozella}, \citenamefont {Proment},\ and\
  \citenamefont {Lvov}}]{Onorato4208}%
  \BibitemOpen
  \bibfield  {author} {\bibinfo {author} {\bibfnamefont {M.}~\bibnamefont
  {Onorato}}, \bibinfo {author} {\bibfnamefont {L.}~\bibnamefont {Vozella}},
  \bibinfo {author} {\bibfnamefont {D.}~\bibnamefont {Proment}}, \ and\
  \bibinfo {author} {\bibfnamefont {Y.~V.}\ \bibnamefont {Lvov}},\ }\href
  {\doibase 10.1073/pnas.1404397112} {\bibfield  {journal} {\bibinfo  {journal}
  {Proc. Natl. Acad. Sci. U.S.A.}\ }\textbf {\bibinfo {volume} {112}},\
  \bibinfo {pages} {4208} (\bibinfo {year} {2015})}\BibitemShut {NoStop}%
\bibitem [{\citenamefont {Lvov}\ and\ \citenamefont
  {Onorato}(2018)}]{PhysRevLett.120.144301}%
  \BibitemOpen
  \bibfield  {author} {\bibinfo {author} {\bibfnamefont {Y.~V.}\ \bibnamefont
  {Lvov}}\ and\ \bibinfo {author} {\bibfnamefont {M.}~\bibnamefont {Onorato}},\
  }\href {\doibase 10.1103/PhysRevLett.120.144301} {\bibfield  {journal}
  {\bibinfo  {journal} {Phys. Rev. Lett.}\ }\textbf {\bibinfo {volume} {120}},\
  \bibinfo {pages} {144301} (\bibinfo {year} {2018})}\BibitemShut {NoStop}%
\bibitem [{\citenamefont {Pistone}\ \emph {et~al.}(2018)\citenamefont
  {Pistone}, \citenamefont {Onorato},\ and\ \citenamefont
  {Chibbaro}}]{0295-5075-121-4-44003}%
  \BibitemOpen
  \bibfield  {author} {\bibinfo {author} {\bibfnamefont {L.}~\bibnamefont
  {Pistone}}, \bibinfo {author} {\bibfnamefont {M.}~\bibnamefont {Onorato}}, \
  and\ \bibinfo {author} {\bibfnamefont {S.}~\bibnamefont {Chibbaro}},\ }\href
  {http://stacks.iop.org/0295-5075/121/i=4/a=44003} {\bibfield  {journal}
  {\bibinfo  {journal} {EPL (Europhysics Letters)}\ }\textbf {\bibinfo {volume}
  {121}},\ \bibinfo {pages} {44003} (\bibinfo {year} {2018})}\BibitemShut
  {NoStop}%
\bibitem [{\citenamefont {Fu}\ \emph {et~al.}(2019{\natexlab{a}})\citenamefont
  {Fu}, \citenamefont {Zhang},\ and\ \citenamefont {Zhao}}]{Our2018}%
  \BibitemOpen
  \bibfield  {author} {\bibinfo {author} {\bibfnamefont {W.}~\bibnamefont
  {Fu}}, \bibinfo {author} {\bibfnamefont {Y.}~\bibnamefont {Zhang}}, \ and\
  \bibinfo {author} {\bibfnamefont {H.}~\bibnamefont {Zhao}},\ }\href {\doibase
  10.1103/PhysRevE.100.010101} {\bibfield  {journal} {\bibinfo  {journal}
  {Phys. Rev. E}\ }\textbf {\bibinfo {volume} {100}},\ \bibinfo {pages}
  {010101(R)} (\bibinfo {year} {2019}{\natexlab{a}})}\BibitemShut {NoStop}%
\bibitem [{\citenamefont {Fu}\ \emph {et~al.}(2019{\natexlab{b}})\citenamefont
  {Fu}, \citenamefont {Zhang},\ and\ \citenamefont {Zhao}}]{Fu_2019}%
  \BibitemOpen
  \bibfield  {author} {\bibinfo {author} {\bibfnamefont {W.}~\bibnamefont
  {Fu}}, \bibinfo {author} {\bibfnamefont {Y.}~\bibnamefont {Zhang}}, \ and\
  \bibinfo {author} {\bibfnamefont {H.}~\bibnamefont {Zhao}},\ }\href {\doibase
  10.1088/1367-2630/ab115a} {\bibfield  {journal} {\bibinfo  {journal} {New J.
  Phys.}\ }\textbf {\bibinfo {volume} {21}},\ \bibinfo {pages} {043009}
  (\bibinfo {year} {2019}{\natexlab{b}})}\BibitemShut {NoStop}%
\bibitem [{\citenamefont {Fu}\ \emph {et~al.}(2019{\natexlab{c}})\citenamefont
  {Fu}, \citenamefont {Zhang},\ and\ \citenamefont
  {Zhao}}]{PhysRevE.100.052102}%
  \BibitemOpen
  \bibfield  {author} {\bibinfo {author} {\bibfnamefont {W.}~\bibnamefont
  {Fu}}, \bibinfo {author} {\bibfnamefont {Y.}~\bibnamefont {Zhang}}, \ and\
  \bibinfo {author} {\bibfnamefont {H.}~\bibnamefont {Zhao}},\ }\href {\doibase
  10.1103/PhysRevE.100.052102} {\bibfield  {journal} {\bibinfo  {journal}
  {Phys. Rev. E}\ }\textbf {\bibinfo {volume} {100}},\ \bibinfo {pages}
  {052102} (\bibinfo {year} {2019}{\natexlab{c}})}\BibitemShut {NoStop}%
\bibitem [{\citenamefont {Wang}\ \emph {et~al.}(2020)\citenamefont {Wang},
  \citenamefont {Fu}, \citenamefont {Zhang},\ and\ \citenamefont
  {Zhao}}]{PhysRevLett.124.186401}%
  \BibitemOpen
  \bibfield  {author} {\bibinfo {author} {\bibfnamefont {Z.}~\bibnamefont
  {Wang}}, \bibinfo {author} {\bibfnamefont {W.}~\bibnamefont {Fu}}, \bibinfo
  {author} {\bibfnamefont {Y.}~\bibnamefont {Zhang}}, \ and\ \bibinfo {author}
  {\bibfnamefont {H.}~\bibnamefont {Zhao}},\ }\href {\doibase
  10.1103/PhysRevLett.124.186401} {\bibfield  {journal} {\bibinfo  {journal}
  {Phys. Rev. Lett.}\ }\textbf {\bibinfo {volume} {124}},\ \bibinfo {pages}
  {186401} (\bibinfo {year} {2020})}\BibitemShut {NoStop}%
\bibitem [{\citenamefont {Mendl}\ \emph {et~al.}(2016)\citenamefont {Mendl},
  \citenamefont {Lu},\ and\ \citenamefont {Lukkarinen}}]{PhysRevE.94.062104}%
  \BibitemOpen
  \bibfield  {author} {\bibinfo {author} {\bibfnamefont {C.~B.}\ \bibnamefont
  {Mendl}}, \bibinfo {author} {\bibfnamefont {J.}~\bibnamefont {Lu}}, \ and\
  \bibinfo {author} {\bibfnamefont {J.}~\bibnamefont {Lukkarinen}},\ }\href
  {\doibase 10.1103/PhysRevE.94.062104} {\bibfield  {journal} {\bibinfo
  {journal} {Phys. Rev. E}\ }\textbf {\bibinfo {volume} {94}},\ \bibinfo
  {pages} {062104} (\bibinfo {year} {2016})}\BibitemShut {NoStop}%
\bibitem [{\citenamefont {Pistone}\ \emph {et~al.}(2019)\citenamefont
  {Pistone}, \citenamefont {Chibbaro}, \citenamefont {Bustamante},
  \citenamefont {L'vov},\ and\ \citenamefont {Onorato}}]{Pistone2018}%
  \BibitemOpen
  \bibfield  {author} {\bibinfo {author} {\bibfnamefont {L.}~\bibnamefont
  {Pistone}}, \bibinfo {author} {\bibfnamefont {S.}~\bibnamefont {Chibbaro}},
  \bibinfo {author} {\bibfnamefont {M.}~\bibnamefont {Bustamante}}, \bibinfo
  {author} {\bibfnamefont {Y.}~\bibnamefont {L'vov}}, \ and\ \bibinfo {author}
  {\bibfnamefont {M.}~\bibnamefont {Onorato}},\ }\href {\doibase
  10.3934/mine.2019.4.672} {\bibfield  {journal} {\bibinfo  {journal} {Math.
  Eng.}\ }\textbf {\bibinfo {volume} {1}},\ \bibinfo {pages} {672} (\bibinfo
  {year} {2019})}\BibitemShut {NoStop}%
\bibitem [{\citenamefont {{Wang}}\ \emph {et~al.}(2020)\citenamefont {{Wang}},
  \citenamefont {{Fu}}, \citenamefont {{Zhang}},\ and\ \citenamefont
  {{Zhao}}}]{2020arXiv200503478W}%
  \BibitemOpen
  \bibfield  {author} {\bibinfo {author} {\bibfnamefont {Z.}~\bibnamefont
  {{Wang}}}, \bibinfo {author} {\bibfnamefont {W.}~\bibnamefont {{Fu}}},
  \bibinfo {author} {\bibfnamefont {Y.}~\bibnamefont {{Zhang}}}, \ and\
  \bibinfo {author} {\bibfnamefont {H.}~\bibnamefont {{Zhao}}},\ }\href@noop {}
  {\bibfield  {journal} {\bibinfo  {journal} {arXiv e-prints}\ ,\ \bibinfo
  {eid} {arXiv:2005.03478}} (\bibinfo {year} {2020})},\ \Eprint
  {http://arxiv.org/abs/2005.03478} {arXiv:2005.03478} \BibitemShut {NoStop}%
\bibitem [{\citenamefont {Toda}(1967)}]{1967Toda}%
  \BibitemOpen
  \bibfield  {author} {\bibinfo {author} {\bibfnamefont {M.}~\bibnamefont
  {Toda}},\ }\href {\doibase 10.1143/JPSJ.22.431} {\bibfield  {journal}
  {\bibinfo  {journal} {J. Phys. Soc. Jpn.}\ }\textbf {\bibinfo {volume}
  {22}},\ \bibinfo {pages} {431} (\bibinfo {year} {1967})}\BibitemShut
  {NoStop}%
\bibitem [{\citenamefont {Ferguson}\ \emph {et~al.}(1982)\citenamefont
  {Ferguson}, \citenamefont {Flaschka},\ and\ \citenamefont
  {McLaughlin}}]{FERGUSON1982157}%
  \BibitemOpen
  \bibfield  {author} {\bibinfo {author} {\bibfnamefont {W.}~\bibnamefont
  {Ferguson}}, \bibinfo {author} {\bibfnamefont {H.}~\bibnamefont {Flaschka}},
  \ and\ \bibinfo {author} {\bibfnamefont {D.}~\bibnamefont {McLaughlin}},\
  }\href {\doibase 10.1016/0021-9991(82)90116-4} {\bibfield  {journal}
  {\bibinfo  {journal} {J. Computat. Phys.}\ }\textbf {\bibinfo {volume}
  {45}},\ \bibinfo {pages} {157 } (\bibinfo {year} {1982})}\BibitemShut
  {NoStop}%
\bibitem [{\citenamefont {Casetti}\ \emph {et~al.}(1997)\citenamefont
  {Casetti}, \citenamefont {Cerruti-Sola}, \citenamefont {Pettini},\ and\
  \citenamefont {Cohen}}]{PhysRevE.55.6566}%
  \BibitemOpen
  \bibfield  {author} {\bibinfo {author} {\bibfnamefont {L.}~\bibnamefont
  {Casetti}}, \bibinfo {author} {\bibfnamefont {M.}~\bibnamefont
  {Cerruti-Sola}}, \bibinfo {author} {\bibfnamefont {M.}~\bibnamefont
  {Pettini}}, \ and\ \bibinfo {author} {\bibfnamefont {E.~G.~D.}\ \bibnamefont
  {Cohen}},\ }\href {\doibase 10.1103/PhysRevE.55.6566} {\bibfield  {journal}
  {\bibinfo  {journal} {Phys. Rev. E}\ }\textbf {\bibinfo {volume} {55}},\
  \bibinfo {pages} {6566} (\bibinfo {year} {1997})}\BibitemShut {NoStop}%
\bibitem [{\citenamefont {Cerruti-Sola}\ \emph {et~al.}(2000)\citenamefont
  {Cerruti-Sola}, \citenamefont {Pettini},\ and\ \citenamefont
  {Cohen}}]{PhysRevE.62.6078}%
  \BibitemOpen
  \bibfield  {author} {\bibinfo {author} {\bibfnamefont {M.}~\bibnamefont
  {Cerruti-Sola}}, \bibinfo {author} {\bibfnamefont {M.}~\bibnamefont
  {Pettini}}, \ and\ \bibinfo {author} {\bibfnamefont {E.~G.~D.}\ \bibnamefont
  {Cohen}},\ }\href {\doibase 10.1103/PhysRevE.62.6078} {\bibfield  {journal}
  {\bibinfo  {journal} {Phys. Rev. E}\ }\textbf {\bibinfo {volume} {62}},\
  \bibinfo {pages} {6078} (\bibinfo {year} {2000})}\BibitemShut {NoStop}%
\bibitem [{\citenamefont {Benettin}\ and\ \citenamefont
  {Ponno}(2011)}]{Benettin2011}%
  \BibitemOpen
  \bibfield  {author} {\bibinfo {author} {\bibfnamefont {G.}~\bibnamefont
  {Benettin}}\ and\ \bibinfo {author} {\bibfnamefont {A.}~\bibnamefont
  {Ponno}},\ }\href {\doibase 10.1007/s10955-011-0277-9} {\bibfield  {journal}
  {\bibinfo  {journal} {J. Stat. Phys.}\ }\textbf {\bibinfo {volume} {144}},\
  \bibinfo {pages} {793} (\bibinfo {year} {2011})}\BibitemShut {NoStop}%
\bibitem [{\citenamefont {Benettin}\ \emph {et~al.}(2013)\citenamefont
  {Benettin}, \citenamefont {Christodoulidi},\ and\ \citenamefont
  {Ponno}}]{Benettin2013}%
  \BibitemOpen
  \bibfield  {author} {\bibinfo {author} {\bibfnamefont {G.}~\bibnamefont
  {Benettin}}, \bibinfo {author} {\bibfnamefont {H.}~\bibnamefont
  {Christodoulidi}}, \ and\ \bibinfo {author} {\bibfnamefont {A.}~\bibnamefont
  {Ponno}},\ }\href {\doibase 10.1007/s10955-013-0760-6} {\bibfield  {journal}
  {\bibinfo  {journal} {J. Stat. Phys.}\ }\textbf {\bibinfo {volume} {152}},\
  \bibinfo {pages} {195} (\bibinfo {year} {2013})}\BibitemShut {NoStop}%
\bibitem [{\citenamefont {Benettin}\ \emph {et~al.}(2018)\citenamefont
  {Benettin}, \citenamefont {Pasquali},\ and\ \citenamefont
  {Ponno}}]{Benettin2018}%
  \BibitemOpen
  \bibfield  {author} {\bibinfo {author} {\bibfnamefont {G.}~\bibnamefont
  {Benettin}}, \bibinfo {author} {\bibfnamefont {S.}~\bibnamefont {Pasquali}},
  \ and\ \bibinfo {author} {\bibfnamefont {A.}~\bibnamefont {Ponno}},\ }\href
  {\doibase 10.1007/s10955-018-2017-x} {\bibfield  {journal} {\bibinfo
  {journal} {J. Stat. Phys.}\ }\textbf {\bibinfo {volume} {171}},\ \bibinfo
  {pages} {521} (\bibinfo {year} {2018})}\BibitemShut {NoStop}%
\bibitem [{not()}]{note1}%
  \BibitemOpen
  \href@noop {} {\bibinfo  {journal} {According to the analysis method
  presented here, it can be easily proved that the harmonic (Toda) chain under
  stress is still a harmonic (Toda) one, only the coefficient of the harmonic
  term in the expansion of the interaction potential will change with the
  stress}\ }\BibitemShut {NoStop}%
\bibitem [{\citenamefont {Matsuda}\ and\ \citenamefont
  {Ishii}(1970)}]{1970Localization}%
  \BibitemOpen
\bibfield  {journal} {  }\bibfield  {author} {\bibinfo {author} {\bibfnamefont
  {H.}~\bibnamefont {Matsuda}}\ and\ \bibinfo {author} {\bibfnamefont
  {K.}~\bibnamefont {Ishii}},\ }\href {\doibase 10.1143/PTPS.45.56} {\bibfield
  {journal} {\bibinfo  {journal} {Prog. Theor. Phys. Supp.}\ }\textbf {\bibinfo
  {volume} {45}},\ \bibinfo {pages} {56} (\bibinfo {year} {1970})}\BibitemShut
  {NoStop}%
\bibitem [{\citenamefont {Livi}\ \emph {et~al.}(1985)\citenamefont {Livi},
  \citenamefont {Pettini}, \citenamefont {Ruffo}, \citenamefont
  {Sparpaglione},\ and\ \citenamefont {Vulpiani}}]{PhysRevA.31.1039}%
  \BibitemOpen
  \bibfield  {author} {\bibinfo {author} {\bibfnamefont {R.}~\bibnamefont
  {Livi}}, \bibinfo {author} {\bibfnamefont {M.}~\bibnamefont {Pettini}},
  \bibinfo {author} {\bibfnamefont {S.}~\bibnamefont {Ruffo}}, \bibinfo
  {author} {\bibfnamefont {M.}~\bibnamefont {Sparpaglione}}, \ and\ \bibinfo
  {author} {\bibfnamefont {A.}~\bibnamefont {Vulpiani}},\ }\href {\doibase
  10.1103/PhysRevA.31.1039} {\bibfield  {journal} {\bibinfo  {journal} {Phys.
  Rev. A}\ }\textbf {\bibinfo {volume} {31}},\ \bibinfo {pages} {1039}
  (\bibinfo {year} {1985})}\BibitemShut {NoStop}%
\bibitem [{\citenamefont {Yoshida}(1990)}]{YOSHIDA1990262}%
  \BibitemOpen
  \bibfield  {author} {\bibinfo {author} {\bibfnamefont {H.}~\bibnamefont
  {Yoshida}},\ }\href {\doibase 10.1016/0375-9601(90)90092-3} {\bibfield
  {journal} {\bibinfo  {journal} {Phys. Lett. A}\ }\textbf {\bibinfo {volume}
  {150}},\ \bibinfo {pages} {262 } (\bibinfo {year} {1990})}\BibitemShut
  {NoStop}%
\bibitem [{\citenamefont {Chen}\ \emph {et~al.}(2014)\citenamefont {Chen},
  \citenamefont {Wang}, \citenamefont {Casati},\ and\ \citenamefont
  {Benenti}}]{PhysRevE.90.032134}%
  \BibitemOpen
  \bibfield  {author} {\bibinfo {author} {\bibfnamefont {S.}~\bibnamefont
  {Chen}}, \bibinfo {author} {\bibfnamefont {J.}~\bibnamefont {Wang}}, \bibinfo
  {author} {\bibfnamefont {G.}~\bibnamefont {Casati}}, \ and\ \bibinfo {author}
  {\bibfnamefont {G.}~\bibnamefont {Benenti}},\ }\href {\doibase
  10.1103/PhysRevE.90.032134} {\bibfield  {journal} {\bibinfo  {journal} {Phys.
  Rev. E}\ }\textbf {\bibinfo {volume} {90}},\ \bibinfo {pages} {032134}
  (\bibinfo {year} {2014})}\BibitemShut {NoStop}%
\bibitem [{\citenamefont {Lepri}(2016)}]{lepri2016thermal}%
  \BibitemOpen
  \bibinfo {editor} {\bibfnamefont {S.}~\bibnamefont {Lepri}},\ ed.,\
  \href@noop {} {\emph {\bibinfo {title} {Thermal Transport in Low Dimensions:
  From Statistical Physics to Nanoscale Heat Transfer}}},\ \bibinfo {series}
  {Lect. Notes Phys.}, Vol.\ \bibinfo {volume} {921}\ (\bibinfo  {publisher}
  {Springer, New York},\ \bibinfo {year} {2016})\BibitemShut {NoStop}%
\bibitem [{\citenamefont {Zhao}\ and\ \citenamefont
  {Wang}(2018)}]{PhysRevE.97.010103}%
  \BibitemOpen
  \bibfield  {author} {\bibinfo {author} {\bibfnamefont {H.}~\bibnamefont
  {Zhao}}\ and\ \bibinfo {author} {\bibfnamefont {W.~-g.}\ \bibnamefont
  {Wang}},\ }\href {\doibase 10.1103/PhysRevE.97.010103} {\bibfield  {journal}
  {\bibinfo  {journal} {Phys. Rev. E}\ }\textbf {\bibinfo {volume} {97}},\
  \bibinfo {pages} {010103(R)} (\bibinfo {year} {2018})}\BibitemShut {NoStop}%
\bibitem [{\citenamefont {Livi}(2020)}]{Livi_2020}%
  \BibitemOpen
  \bibfield  {author} {\bibinfo {author} {\bibfnamefont {R.}~\bibnamefont
  {Livi}},\ }\href {\doibase 10.1088/1742-5468/ab7125} {\bibfield  {journal}
  {\bibinfo  {journal} {J. Stat. Mech. Theory Exp.}\ }\textbf {\bibinfo
  {volume} {2020}},\ \bibinfo {pages} {034001} (\bibinfo {year}
  {2020})}\BibitemShut {NoStop}%
\bibitem [{\citenamefont {Lepri}\ \emph {et~al.}(2020)\citenamefont {Lepri},
  \citenamefont {Livi},\ and\ \citenamefont {Politi}}]{PhysRevLett.125.040604}%
  \BibitemOpen
  \bibfield  {author} {\bibinfo {author} {\bibfnamefont {S.}~\bibnamefont
  {Lepri}}, \bibinfo {author} {\bibfnamefont {R.}~\bibnamefont {Livi}}, \ and\
  \bibinfo {author} {\bibfnamefont {A.}~\bibnamefont {Politi}},\ }\href
  {\doibase 10.1103/PhysRevLett.125.040604} {\bibfield  {journal} {\bibinfo
  {journal} {Phys. Rev. Lett.}\ }\textbf {\bibinfo {volume} {125}},\ \bibinfo
  {pages} {040604} (\bibinfo {year} {2020})}\BibitemShut {NoStop}%
\end{thebibliography}%

\end{document}